\documentclass[11pt]{article}
\usepackage[dvips]{epsfig}
\usepackage{latexsym}
\usepackage{graphicx}
\usepackage{psfrag}
\usepackage[bf,footnotesize]{caption}   
\begin{document}
\title{$\quad$ $\quad$ $\quad$ $\quad$ $\quad$  Beyond the semi-classical description
$\quad$ $\quad$ $\quad$ $\quad$ $\quad$  of black hole 
evaporation\footnote{An earlier version of this work was
 published in Proceedings\cite{Parentani:2002mb}
in 2002. We postponed submitting it to the arXiv in the hope 
of improving 
the evaluation of radiative corrections.
We have recently modified the text, corrected some mistakes, and
added remarks
on this difficult point which still needs work --
that hopefully someone will take over.}}

\author{{\bf{Renaud Parentani}}\footnote{E-mail: Renaud.Parentani@th.u-psud.fr}\\
Laboratoire de 
Physique Th\'eorique,\\
{\it UMR CNRS 8627,}\\
 Universit\'e Paris-Sud,\\ 
 91405 Orsay Cedex, France.} 

\maketitle
\vskip 2.5 cm

\noindent {\bf Abstract:} 

In the semi-classical treatment, i.e. in 
a classical black hole geometry, Hawking quanta emerge from
trans-Planckian configurations because of scale invariance.
There is indeed no scale to stop the blue-shifting effect 
encountered in the backward propagation towards the event horizon. 
On the contrary, when taking into account the gravitational 
interactions neglected in the semi-classical treatment, a 
UV scale
stopping the blue-shift 
could be dynamically engendered.
To show that this is the case,
we use 
a non-perturbative treatment based on 
the large-$N$ limit, where $N$ is the number of matter fields.  
In this limit, the semi-classical treatment is the leading contribution.
Non-linear gravitational effects appear in the next orders
and in the first of these, the effects 
are governed by the two-point correlation function 
of the energy-momentum tensor evaluated in the vacuum.
Taking this correlator into account,
backward propagated modes are dissipated 
at a distance from the horizon
$\propto G\kappa$ when measured in a freely falling frame. ($G$ is Newton's constant and $\kappa$ the surface gravity.)
This result 
can be also obtained by considering light 
propagation in a stochastic ensemble of metrics whose 
fluctuations 
are determined 
by the above correlator.


\vskip 2 cm

\newcommand{\ket}[1] {\mbox{$ \vert #1 \rangle $}}
\newcommand{\bra}[1] {\mbox{$ \langle #1 \vert $}}
\newcommand{\bkn}[1] {\mbox{$ < #1 > $}}
\newcommand{\bk}[1] {\mbox{$ \langle #1 \rangle $}}
\newcommand{\scal}[2]{\mbox{$ < #1 \vert #2 > $}}
\newcommand{\expect}[3] {\mbox{$ \bra{#1} #2 \ket{#3} $}}
\newcommand{\ki}{\mbox{$ \ket{\psi_i} $}}
\newcommand{\bi}{\mbox{$ \bra{\psi_i} $}}
\newcommand{\p} \prime
\newcommand{\e} \epsilon
\newcommand{\la} \lambda
\newcommand{\om} \omega   
\newcommand{\Om} \Omega
\newcommand{\cc}{\mbox{$\cal C $}}
\newcommand{\w} {\hbox{ weak }}
\newcommand{\al} \alpha
\newcommand{\bt} \beta
\newcommand{\be} {\begin{equation}}
\newcommand{\ee} {\end{equation}}
\newcommand{\ba} {\begin{eqnarray}}
\newcommand{\ea} {\end{eqnarray}}
\def\lrD{\mathrel{{\cal D}\kern-1.em\raise1.75ex\hbox{$\leftrightarrow$}}}
\def\lr #1{\mathrel{#1\kern-1.25em\raise1.75ex\hbox{$\leftrightarrow$}}}
\def\ie{{\emph{i.e.}}}
\def\eg{{\emph{e.g.}}}
\def\cf{{\emph{c.f.}}}
\def\d{{\mathrm{d}}}

\newpage

\section{Introduction}

In his original derivation~\cite{Hawk}, Hawking considered the
propagation of quantum radiation in a {\it fixed} background metric, that
of a collapsing star.  This means that the metric is once for all
determined by the energy of the collapsing star. It is therefore
unaffected by the quantum processes under examination.  In this
approximation, the radiation field satisfies a linear equation (in the
absence of matter interactions). One then finds that 
in-falling and outgoing field configurations are completely {\it uncorrelated}
near the black hole horizon.\footnote{In this paper 
we consider only $s$-waves and neglect the residual 
potential barrier which partially back-scatter some Hawking quanta
thereby inducing some correlations. These however play no role in what follows
and shall be ignored.
In the next equations, 
 $v$ and $u$ are radial advanced and retarded
null coordinates.
In the Schwarzschild geometry, they are given by 
$v=t+r^*, u=t-r^*$ where $r^*= r + 2M \ln (r /2M -1 )$ is the
tortoise coordinate~\cite{MTW}.}  
This is explicitized by the fact that the connected part of the 
two-point correlation function
of $T_{uu}$ and $T_{vv}$, the energy-momentum tensor of outgoing 
and infalling configurations,
\be
\langle T_{vv}(x) \;T_{uu}(x')\rangle_C = 
\langle T_{vv}(x) \;T_{uu}(x')\rangle - 
\langle T_{vv}(x)\rangle \; \langle T_{uu}(x')\rangle \, , 
\label{8TT}
\ee 
vanishes in the (Unruh) 
vacuum.
Nevertheless, Hawking radiation is pair creation.
This is perfectly consistent with eq. (\ref{8TT})
since the pairs are composed of two outgoing quanta, 
one of each side of the event horizon.  
The external ones form the asymptotic flux whereas their partners
propagate 
towards the singularity at $r=0$.  
Upon tracing over these inner configurations, one gets an
 incoherent flux described by a thermal density 
matrix\cite{GO}.\footnote{
The non-vanishing character 
of the correlation function 
$\langle T_{uu} \;T_{uu}\rangle_C$ 
across the horizon demonstrates that 
the partners of Hawking quanta are inside outgoing 
configurations. 
An explicite calculation gives\cite{CW,MP2}
\be
\langle T_{uu}(u) \;T_{uu}(u'_L)\rangle_C \propto 
\vert u-u'_L + i \pi /\kappa \vert^{-4} \, ,\label{uucor}\ee
 where $\kappa u_L = \ln(\kappa U_K)$ is a null coordinate
 for the inside configurations. $U_K$ is the usual Kruskal retarded time; it 
 vanishes on the horizon, and 
is positive inside the hole. $\kappa=1/4M$ is the surface gravity. (It fixes 
Hawking temperature $T_{H}= \kappa/ 2 \pi$. 
We work in Planck units: $c=\hbar = M_{Planck} =1$.)
The smooth maximum of this two-point function 
for opposite points, i.e. $ u =u'_L$ or
$U_K = - U_K'$, is a direct consequence of the fact that each
pair is indeed composed of two outgoing quanta leaving on either side.
}

From this fixed background description, one may go one step further 
by performing a mean field approximation, i.e. by including 
the metric change determined by
Einstein's equations driven by the expectation value
$\langle T_{\mu \nu} \rangle$.
One then finds that this expectation value is 
{\it regular} \cite{Bardeen,PP,massar,GO}.
This 
guarantees that the black hole will adiabatically
evaporate while keeping the regularity of the near horizon geometry.
This regularity in turn implies that the infalling and 
outgoing configurations will stay {uncorrelated}.
Therefore,  the correlation function (\ref{8TT})
still vanishes
in the semi-classical treatment.

This adiabatic evolution would provide a reliable starting point for 
including perturbatively radiative corrections were
another feature of black hole physics not present. Namely, the
field configurations giving rise to Hawking quanta possess
arbitrary high (trans-Planckian) frequencies near the 
horizon \cite{jaco,jaco93,EMP,GO}. When 
measured by free falling (FF) observers  at $r$, the frequency of 
an outgoing quantum of asymptotic energy $\la$ grows as
\be
\Om \propto {\la \over 1 - 2M /r } \, .
\label{dopef}
\ee
This implies
that 
wave packets centered along the null outgoing geodesic $u$
had FF frequency growing as $\Om \propto\la e^{\kappa u}$ 
when they emerged from the collapsing star.  
Unlike processes
characterized by a typical energy scale, the relation $\Om \propto \la
e^{\kappa u}$ shows that black hole evaporation rests, in this
scenario, on arbitrary high frequencies.  
This conclusion drawn from the analysis of wave packets is confirmed by the
study of the non-diagonal matrix elements of $T_{\mu \nu}$. These 
characterize the fluctuations of the flux around its mean value.  As shown
in \cite{MP2,EMP,GO}, unlike 
the expectation value (the diagonal part)
which is regular and of the order of $M^{-4}$, these matrix elements
are 
{\it singular} on the horizon, \ie, their Fourier
content is characterized by FF frequencies which grow according to
eq. (\ref{dopef}).
 
This growth is not an artifact of a bad coordinate choice. Indeed,
 as emphasized by 't~Hooft~\cite{THooft}, and explained below in Section 3,
gravitational interactions between the configurations giving rise to
Hawking quanta and in-falling quanta %
also grow according to eq. (\ref{dopef}).
This questions the validity of the semi-classical treatment and 
the vanishing of
eq. (\ref{8TT}).\footnote{In section 9 of the 
review\cite{THooft}, one reads
``Any decomposition of Hilbert space in terms of mutually
non-interacting field quanta will be hopelessly inadequate in this
(near horizon) region.''}

In questioning the validity of the semi-classical description, two issues
should be distinguished, see {\it e.g.} section 3.7 in \cite{GO}.  First, there
is the question of low frequency $O(\kappa)$ 
modifications of the asymptotic properties of Hawking radiation, and secondly, 
that of high frequency
modifications of the near horizon physics.  
Since all thermo-dynamical
reasonings indicate that the asymptotic properties (namely thermality
governed by $\kappa$ and stationarity) should be preserved, 
the problem is to reconcile 
this robustness with the radical change of
the near horizon physics which is needed to 
tame the growth 
of gravitational 
interactions. 
This is not an easy problem: Indeed, the  
perturbative
analysis of near horizon interactions performed in ~\cite{hawkfr}
leads to back-reaction effects which grow like $\Om$ in eq. (\ref{dopef}).
This threatens the stationarity of the flux and therefore
questions the {\it choice} of the treatment which is adequate
to go beyond
the semi-classical approximation.

As a first step towards a full quantum gravitational treatment, 
inspired by~\cite{THooft}-\cite{Barr}, 
we proposed\cite{hawkm,Parentani:2001tg,Parentani:2002mb} 
a non-perturbative treatment of the  interactions occuring in the
FF {vacuum}
based on $\langle T_{vv} 
T_{vv} 
\rangle_C$, the 
correlation function  of infalling configurations. 
As discussed in more detail in what follows,
this treatment emerges in a large $N$ limit,
where $N$ is the number of copies of the quantum field.  
In physical terms, in this limit, in-falling
configurations act as an environment for the outgoing quanta
and their 
gravitational interactions express themselves in
terms of a stochastic ensemble of metric fluctuations.  
The statistical properties of the latter are determined by the fluctuations
of the $N$ fields in the
FF vacuum. Their main effect 
is to
dissipate the outgoing modes traced backwards
towards the horizon at the locus 
where their FF frequency  (\ref{dopef}) reaches $1/\bar \sigma$, 
i.e., at  $r = 2M + \bar \sigma$ for modes of Killing frequency equal to $\kappa$. 

In our model, the gravitationally induced 
 length-scale\footnote{\it \label{proper}
{\underline{Added comment.}}
When presented in meetings, this result was 
skeptically received 
on the basis that $\bar \sigma$ 
is much shorter than the Planck length for macroscopic black holes. Indeed one has
$\bar \sigma \propto L_{Planck}^2 /G M_{bh}$. 
However it should be pointed out 
that $\bar \sigma$, which is the proper distance between 
$2M$ and $2M + \bar \sigma$ measured 
in a FF frame,
actually corresponds to a proper distance of the order of $L_{Planck}$
when measured along surfaces of constant (Killing) time.
It should be also recalled that $r$ is  
defined by the square root of the area
of 2-spheres$/4\pi$, and is therefore coordinate invariant.
On the contrary, the notion of distance from the horizon
depends on the choice of the time slices used. In this sense
our effective propagation law does {not} singles out a {\rm prefered frame}.
This should be compared with 
the procedure involved when using an a priori given dispersion relation,
see \cite{dumbunruh}-\cite{Jacobson:2007jx}. 
I am grateful to Ted Jacobson for discussions
on this.}
is
\be
\bar \sigma 
\propto 
{\kappa \over M_{Planck}^2} \sqrt{ N \ln(\Lambda/\kappa)}\, . 
\label{news}
\ee
In the above $\Lambda$ is a high frequency cut-off
whose value requires further study to be fixed. 
It is
important to notice that in spite of these dissipative effects, 
the gravitational interactions do not significantly
affect the asymptotic properties of Hawking radiation.
As shown in \cite{Barr}, the asymptotic corrections scale indeed as
 $(\kappa \bar \sigma)^2$, hence they are negligible
for macroscopic black holes.
It should also be noticed that
the dissipative effects 
break the 
2D local~\cite{jaco} Lorentz invariance.\footnote{In the vicinity of a 
black hole horizon, free propagation of $s$-waves
is governed by
a 2D Lorentz (and scale) invariance in the $u, v$ plane, see Section 2.
When including radiative corrections, the dressed propagator looses 
this property. 
This is similar to the fact that 
the self-energy of an electron immersed
in a thermal bath of photons is not Lorentz invariant either\cite{APV}. 
In both cases,
the integrands governing loop corrections are not Lorentz invariant, more on this 
in Section 5.}

An unsolved question concerns the range of validity of our treatment.
This is a 
complicated question whose final answer requires a better
understanding of quantum gravity.  Let us nevertheless make some remarks.  
First, this question closely follows that concerning the validity of the
semi-classical treatment.\footnote{The validity of the semi-classical treatment has
been often questioned in rather general terms. However, a significant
answer requires to find physical quantities (\ie, matrix elements of
observables) which are {\it incorrectly} evaluated in this treatment {\it and}
to propose improved expressions for the same quantities in order to
see the discrepancy. We shall provide an explicit exemple
in Section 4.}
Secondly, our analysis indicates that the
semi-classical treatment fails {before} our treatment.  `Before'
should be understood radially, given the blue shift effect (\ref{dopef})
encountered
in the backward propagation of outgoing configurations.
What emerges is a kind of Russian doll
structure in which quantum gravity progressively dominates the physics.
 Far
away from the hole ($r-2M
\gg 2M$) one has outgoing thermal (on shell) radiation.  In a first
intermediate regime ($\bar \sigma \ll r-2M \ll 2M$) the propagation of
outgoing modes is 
still governed by the d'Alembertian but observers at fixed
$r$ and free falling ones perceive quanta differently (in that the
locally defined
FF vacuum no longer agrees with the Killing (Boulware) vacuum).  
It is in this regime (well described by the semi-classical treatment) 
that Hawking radiation gets established,
see eqs. (76-83) in \cite{MP2}. This description
based on modes stops to be valid when reaching a Jacobson's time-like
boundary~\cite{jaco93}, at $r \simeq  2M + \bar \sigma$, when outgoing modes
get progressively entangled to the infalling configurations, thereby
loosing their `mode' quality. The principle aim of this paper is to
analyze this transitory regime.  Deeper in $r$, one has some
unknown regime governed by Planckian physics.  This physics presumably
also occurs around us but stays well hidden inside its Planckian husk
in the absence of a good microscope.

\vskip .3 truecm 

Below we discuss two aspects.
We first explain why we adopted a treatment based on the 
large $N$ limit, and then we further discuss the relation between 
this treatment and the dispersive physics involved in acoustic black holes. 
These two subsections can be skipped in a first reading.

\subsection{
The choice of the treatment: the large-$N$ limit.}


In order 
to compute/estimate
 quantum gravitational corrections to Hawking radiation
 in the absence of  
a theory of quantum gravity,
one should 
adopt an approximative treatment 
allowing to compute radiative corrections 
to some physical quantities.
In this paper, 
we have chosen a statistical treatment based on a large-$N$ limit,
where $N$ is the number of copies of the quantum 
matter field.  The reason for
this choice are the following.

First, the semi-classical treatment {\it is} the leading contribution
in the large-$N$ limit. This is most simply understood in a path integral 
approach~\cite{hh}: By performing a saddle point approximation
in the evaluation the one-loop effective action, one verifies that the 
location of the saddle is determined by the semi-classical 
Einstein equations.
Hence, the semi-classical treatment is the mean field
approximation (Hartree) in which the $N$ copies of the 
radiation field propagate in a self-consistent classical geometry:
a solution of Einstein equations driven by $N$ times $\bk{T_{\mu\nu}}$,
the mean value of the energy-momentum tensor of one field.
This result can be also understood from a diagrammatic point of view
in the following way. One first verifies that the expectation value of 
any observable can be expanded as a {\it double} series 
in $G$ and $N$ in which the power of $N$ is always smaller or equal
to that of $G$. One then verifies that the semi-classical value of 
this observable corresponds to the resummed series containing 
all terms governed only by the one-point 
function $\bk{T_{\mu\nu}}$. Moreover, all these graphs 
are 1-particle reducible and their weight is a positive power of $GN$.
 This analysis furnishes an alternative proof that 
the semi-classical treatment is the leading contribution
and that it corresponds to a mean field treatment, see the Appendix
for more detail. 

This diagrammatic analysis naturally leads to inquire about the next series. 
Upon having first summed up the leading series in powers of $GN$,
one encounters a next series containing positive powers of $G^2 N$. 
This second series is governed by the two-point function 
$\langle T_{\mu \nu}(x) \; T_{\alpha \beta}(x')\rangle_C$, the
`specific heat' of the radiation field. 
The second reason of having chosen the large $N$ limit is
that non-perturbative effects can be obtained by resumming this second
series\cite{hawkm}. 

The third reason which has led us to choose a treatment 
based on a statistical basis 
arises from the trans-Planckian problem. 
Indeed, as previously discussed, 
the unbounded growth of frequencies
encountered near the event horizon
seems to invalidate
perturbative treatments of radiative corrections, see \cite{hawkfr}
for an exemple of the divergences encountered.

Let us briefly explain what is the nature of the trans-Planckian problem
and why radiative corrections induced by quantum gravity
might give some important effects when applied to Hawking
radiation (or to quantum effects induced by the presence of an event
horizon).  
When studying the {\it origin} of Hawking quanta one faces
a difficulty which is specific to horizon physics: The
configurations giving rise to Hawking quanta are characterized by
ultra-high frequencies when measured by infalling observers near the
horizon~\cite{Unruh81,jaco,THooft,MP2}.  Indeed, in the semi-classical
treatment, the use of free fields propagating in a classical background
leads to {unbounded} frequencies as a direct consequence of the
structure of the outgoing null geodesics near the horizon.  Therefore,
{\it any} ultra-violet scale which would signal the breakdown of the
semi-classical treatment will be {\it inevitably} reached. 

This reasoning is nicely illustrated by considering sound propagation in an
acoustic geometry which possesses a horizon: One first finds 
that the propagation is dramatically modified with respect to the standard 
propagation (governed by the d'Alembertian) when the FF frequency $\Omega$
 reaches $\Omega_c$, the UV scale (the inverse inter-atomic
distance). In particular, for all non-linear dispersion 
relations, one finds  that the focussing effectively 
stops when this new scale is reached. 
Secondly, this dramatic
modification of the near horizon propagation leaves no imprint on 
the asymptotic properties of Hawking radiation when the inverse inter-atomic distance is well separated~\cite{dumbunruh,bmps} from $\kappa$, 
the surface gravity of the hole.

Our aim is to show that similar results are obtained when  
computing non-perturbatively the gravitational effects driven by 
the connected two-point function 
$\langle T_{\mu \nu} \; T_{\alpha \beta}\rangle_C$.
We shall find that the trans-Planckian 
correlations which existed in the semi-classical treatment 
are washed out when the $r-2M$ reaches $\bar \sigma$,
 the {\it new} length scale which plays the role 
of the inter-atomic distance. 
Moreover, as already mentioned, this washing-out mechanism leaves 
the asymptotic properties of Hawking radiation unaffected: 
the thermal flux receives corrections which
scale like $(\kappa \bar \sigma)^2$ and which are therefore negligible
for large black holes.


\subsection{The lesson from 
acoustic black hole physics.}

For the interested reader, we further discuss the 
relationships between our approach and the physics of sonic black holes.
As just explained, the appealing feature of these models is to provide both 
a simple explanation (in terms of adiabaticity 
which essentially follows from scale separation
$\Om_c \gg \kappa$) for the robustness of the asymptotic
properties of the flux, and a simple physical reason (a modified
dispersion relation) which eradicates the ultra-high frequencies.
(It should be pointed out that a similar trans-Planckian problem
arises in inflationary models when studying the origin of 
the spectrum of primordial energy density fluctuations~\cite{tedjap,MB,N}.
In that case as well, scale separation and regularity
of the metric are sufficient conditions to guarantee that 
the properties of the spectrum are unmodified~\cite{np}.)

Besides the robustness of the IR properties, the main
outcome of these considerations is that a {new universality}
has emerged: for {\it all} dispersion relations but the linear one, 
the blue shifting effect stops. Therefore, the never ending
blue shifting effect obtained by using the linear
(scale-less and non-dispersive) relation $\Omega = p$ now appears
as an isolated and unstable behaviour. Thus, 
instead of asking:

{\it{\bf is Hawking radiation robust against
modifying the dispersion relation ?}}
 
\noindent
we are led to question\cite{dublin}:

{\it {\bf  is $\Omega = p$ robust against radiative corrections ? }}

\noindent
or it is simply an artifact of free field theory ?

These considerations suggest that quantum gravity should
engender a new UV scale when evaluating 
radiative corrections in a black hole geometry.
This new scale would then break the scale invariance of 
free field propagation and prevent the appearance of trans-Planckian 
frequencies. 
To verify this conjecture, one must determine the physical effects 
induced by the {non-linearities} engendered 
by gravitational interactions.
When this is done, one can 
make contact with 
sound wave physics\cite{rmed} by determining
how phenomenologically describe by an effective 
linear equation (\ie,  a non-trivial dispersion relation)
the dissipative/dispersive effects induced by non-linearities.

In this paper we shall implement the second question 
by computing how the radiative corrections driven by
$\bk{T_{vv} T_{vv}}$ modify the two-point function
of out-going configurations. We shall see that $\Omega = p$ 
is not robust in that it provides a good approximation
of the effective propagation law only when the FF frequency
$\Omega$ obeys $\Omega \ll 1/\bar \sigma$ where $\bar \sigma$
is given in eq. (\ref{news}).


\newpage

\section{The Model}

For simplicity, we shall consider only $s$-waves propagating in
spherically symmetric space times.  For definiteness, the
background metric is taken to be that resulting from the collapse of a null
shell of mass $M_0$ which propagates along the null ray $v=0$.  
Inside the shell,
for $v<0$, the geometry is Minkowski and described by 
\be
ds^2 = - ( 1 - {2M(v) \over r}) dv^2  + 2 dvdr + r^2(d\theta^2 + \sin^2 
\theta d\phi^2) \, .
\label{two}
\ee 
with $M=0$.  Outside the shell, the metric is Schwarzschild 
and given by eq. (\ref{two}) with $M(v)=M_0$.  As we shall see,
this choice of the collapsing metric will have no
influence in what follows since 
the near horizon vacuum interactions are stationary.

\begin{figure}[ht] 
\epsfxsize=6.5cm
\psfrag{scryp}{${\cal J}^+$}
\psfrag{phip}{$\phi_+$}
\psfrag{scrym}{${\cal J}^-$}
\psfrag{phim}{$\phi_-$}
\psfrag{veqo}{$v=0$}
\psfrag{reqo}{$r=0$}
\psfrag{singularity}{singularity}
\centerline{{\epsfbox{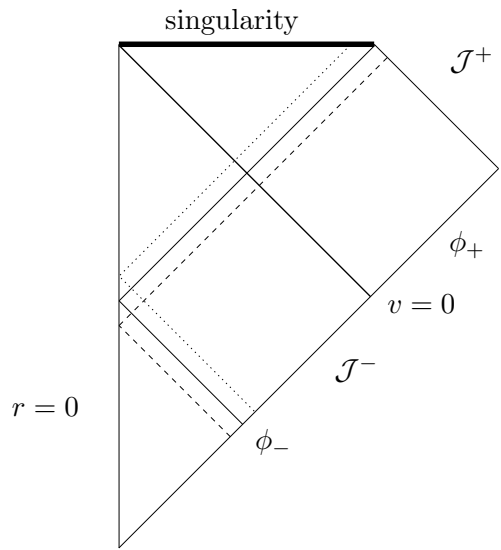}}}
\caption{{\bf Penrose diagram of the background geometry.}
The light-like infalling shell propagates along $v=0$. 
The other continuous line 
emerging from ${\cal J}^-$ is $v=v_H$, the radial light ray which 
forms the event horizon after having bounced on $r=0$.
The dashed line represents a characteristic of the configurations
$\phi_-$ which are responsible for Hawking quanta. The dotted line 
represents a partner's characteristic. 
For quanta reaching ${\cal J}^+$ at late
$u$, both of these characteristics are extremely close to $v_H$,
see eq. (\ref{exponrel}). The configurations
$\phi_+$ have support for $v>0$ and are always infalling. 
In this paper, we study the interactions between 
$\phi_-$ and $\phi_+$ which occur outside the star,
in the near horizon empty region,  when the (initial) state of 
$\phi_+$ and $\phi_-$ is vacuum.}
\label{Fig1}
\end{figure}

To identify the degrees of freedom involved in these interactions, we
first analyze the global properties of radial modes,
when working in the geometric optic approximation, i.e. 
when working with $\partial_u \partial_v \phi =0$. 
(In the exact d'Alembertian, see eq. (\ref{4Ddal}), 
there is a potential around $r=3M$ which induces partial reflection, 
a phenomenon irrelevant for our purposes.)
The ingoing massless waves fall into two classes according to their 
support on ${\cal{J}}^-$, the light-like past infinity.  
The waves in the first class have support only for $v<0$,
inside the shell, and will be noted $\phi_-$.  They
propagate inward in the flat geometry till $r=0$ where they bounce off
and become outgoing configurations, see Figure 1.  
The relationship between the
value of $u$ of the geodesic which originates from $v$ on
${\cal{J}}^-$ is~\cite{Barr}: 
\be
V(u)= -4M ( 1 + e^{-ku})\, ,
\ee
 The first
class is thus divided in two subsectors: For $v<-4M$, the reflected
waves cross the in-falling shell with $r>2M$ and reach the asymptotic
region whereas those for $0>v>-4M$ cross the shell with $r<2M$ and
propagate in the trapped region till the singularity.  The separating
light ray $v_H=-4M$ becomes the future horizon $u=\infty$ after
bouncing off at $r=0$.  The configurations which form the second class
live outside the shell, have support only for $v>0$ and 
are noted $\phi_+$. They propagate in the static Schwarzschild geometry,
are always in-falling and cross the horizon towards the singularity.

In Hawking's derivation of black hole radiation, the field operator 
obeys the d'Alembert equation. Hence the above classical 
properties 
apply: The configurations for $v<v_H$ give rise to
the asymptotic quanta, those for $v_H<v<0$ to their partners~\cite{MP2}
whereas the configurations described by
$\phi_+$ play no role in the asymptotic radiation. 
This 
follows from the asymptotic ($\kappa u \gg 1$)
behaviour of the relation $V(u)$
\be
V(u)-v_H \propto e^{-\kappa  u} \, .
\label{exponrel}
\ee
As shown in \cite{Hawk}, this exponential is responsible for the
thermal radiation at temperature $\kappa / 2 \pi$. 
It also shows that Hawking quanta emerge from trans-Planckian 
frequencies on ${\cal J^-}$ since $\om dV = \la du$
(where $ \om = i\partial_v $ on ${\cal{J}}^-$ and
$ \la = i\partial_u $ on ${\cal{J}}^+$) 
gives $\om \propto \la e^{\kappa u}$. 
Finally it fixes the correlations between the asymptotic quanta and their
partners, see eq. (\ref{uucor}). These follow from the fact that, on ${\cal{J}}^-$ and in the
vacuum, the rescaled field $\phi = \sqrt{4 \pi r^2}
\chi$ (where $\chi$ is the 4D $s$-wave) satisfies 
\be
\langle \phi(v) \; \phi(v') \rangle = \int_0^{\infty}{ \d\om \over 4\pi \om } 
e^{-i\om(v-v')}
= - {1 \over 4\pi } \ln( v - v' - i \epsilon) + \mathrm{constant}. \,  
\label{phi2p}
\ee

Since this equation is valid for all $v, v'$ one might think that
there also exist correlations between $\phi_-$ and $\phi_+$. However,
for late Hawking quanta, they effectively vanish since these quanta
and their partners emerge from configurations which are 
localized extremely close to $v_H$ as indicated in eq. (\ref{exponrel}).  
For a description of the other 
properties of Hawking
radiation, we refer to the review \cite{GO}. 


In brief, in the absence of gravitational interactions, 
$\phi_-$ and $\phi_+$ are effectively two independent fields. By
independent we mean that by sending quanta described by wave packets
built with $\phi_+$, there is {\it no} induced emission of Hawking 
quanta. Indeed, in order to get induced emission~\cite{wald} at time $u$, 
one should send $\phi_-$ quanta localized close to $v_H$ as indicated
in eq. (\ref{exponrel}) and correspondingly
characterized by high frequencies $\om \propto \la e^{\kappa u} \gg \kappa$.

Let us now analyze more closely how these properties translate 
in Fock space. 
When evaluated in the background eq. (\ref{two}), 
the action of $\phi$ is
\be
S^\phi_{g} = -
\int  \! \d v \; \d r \left[ \partial_v \phi \; \partial_r \phi + 
\left(1 - {2M \over r} \right){ {(\partial_r \phi )^2 \over 2} } \right] \, ,
\label{actS}
\ee
with $M(v)=0$ for $v<0$ and $M(v)=M_0$ for $v>0$.  Being interested in
the near horizon physics, we have dropped the potential term of
$s$-waves, $(2M_0/r^3) \phi ^2$, since it does not affect the near
horizon propagation.  This can be seen by using the double null
coordinate system $u=v-2r^*, v$. Using them, the 4D d'Alembertian
reads
\be
\left[
\partial_u \partial_v - \left(1-{2M_0 \over r}\right)
\left( {l(l+1) \over r^2} + {2M_0 \over r^3}\right)
\right] \phi_l =0  \, 
\label{4Ddal}
\ee
where $\phi_l$ is the rescaled mode of angular momentum $l$.  Thus, as
emphasized in \cite{Verl3}, the propagation of waves (at fixed angular
momentum and even for an arbitrary mass) effectively obeys a 2D
conformal invariance in the near horizon geometry.\footnote{This
invariance 
leads directly to the trans-Planckian problem: the {\it steady} production
rate of outgoing quanta arises from an integral over in-frequencies
$\omega$ whose measure is that of a 2D massless field.  Explicitly one
obtains that the thermal distribution is multiplied by
$\d\omega/\omega = \kappa \d u$ since $\omega \propto e^{\kappa u}$,
see \cite{GO,hawkfr}.} 
This is confirmed by the fact that, classically, 
the trace of 2D part of $T_{\mu \nu}$ vanishes independently
of the equations of motion. 
Thus, in our model for $s$-waves, $T_{\mu \nu}$ has only two {\it q}-number
components, $T_{vv}=(\partial_v \phi)^2$ and $T_{uu}=(\partial_u\phi)^2$.

The 2D conformal
invariance also implies that the Fock space is composed of tensorial
products of in-falling states (on which $\phi_+$ acts) and outgoing
states.  In a Schroedinger language this means that an initially
factorized state (\ie, $\ket{\Psi}=\ket{\Psi_+} \otimes \ket{\Psi_-}$)
remains factorized when its evolution is governed by
eq. (\ref{actS}). This factorizability 
explains the absence of induced emission
when adding $\phi_+$ quanta.  
In a Heisenberg language, 
it tells us that any matrix element of $\phi$ is a
combination of matrix elements of $\phi_-$ and $\phi_+$ evaluated
separately. This implies in particular that
the connected part of the two-point correlation eq. (\ref{8TT})
identically vanishes for all factorized states. 
This applies to the ``Unruh'' vacuum, 
the state describing Hawking radiation.
Physically, the vanishing of eq. (\ref{8TT}) means that the {\it
fluctuations} of $ T_{vv}$ and $T_{uu}$ around their mean values 
are completely uncorrelated. 
It should 
be clear that this absence of quantum correlations is precisely what 
is contested by t'Hooft, see footnote 3.

Finally, we mention that, in spite of this absence of correlations, 
 the mean value of $T_{vv}$ and $T_{uu}$ are related to each
other by energy conservation through the 2D trace anomaly~\cite{cfu}.
However, this third component of $T_{\mu \nu}$ does not fluctuate:
it is a $c$-number. 
Hence it 
cannot play any role in the
gravitational interactions between the field operators ${\bf \phi_-}$ 
and ${\bf \phi_+}$.

\section{The gravitational interactions between ${\bf \phi_-}$ 
and ${\bf \phi_+}$}

The aim of this Section is to obtain the dominant part 
of the action governing the
gravitational interactions between $\phi_-$ and $\phi_+$.  
In the next Sections, we shall study the consequences of these
interactions with particular emphasis on the correlations they 
induce. 

The generating functional governing our matter-gravity system is
\be
Z= \int \! {\cal{D}} \phi
\,\, {\cal{D}}h \,
\, e^{i[S^\phi_{g+h}+ S_{h, g}]} \, .
\label{Z1}
\ee
$h$ is the change of the metric with respect to the
background $g$ discussed above and $S_{h, g}$ is the Einstein-Hilbert action.
(We only consider metric fluctuations in the Ricci flat region
outside the infalling matter forming the hole.)
$S^\phi_{g+h}$ is the action of $\phi$ propagating in the 
fluctuating geometry $g+h$.

When the metric fluctuations are spherically symmetric, $h$ can be
characterized by two functions: $\psi, \mu$. Moreover both
are completely determined by the energy-momentum tensor of $\phi$.
This is like the longitudinal part of the 
electro-magnetic field which is constrained to follow charge density 
fluctuations, by Gauss' law.  
The line element in the fluctuating metric can be written 
as~\cite{Barr}~\footnote{
This line element differs from that used by Bardeen~\cite{Bardeen}:
\[
\d s^2 =
e^{\psi}\left[
- e^{\psi}\left(1 - {2M_0 + 2\mu_B \over r } \right) \d v^2 + 2 \d v\; \d r
\right] +
r^2 \d\Omega^2_2. 
\]
The $\psi$ function is the same whereas, to first order in $\psi$ and
$\mu_B$, the mass fluctuation 
$\mu= \mu_B - \psi (r-2M_0)/2$.  The usefulness of our choice
is that $\psi$ no longer affects the null geodesics.  When expressing $T_{\mu \nu}$
in terms of the null fluxes $T_{vv},T_{uu}$, 
Einstein's equations give \be\partial_v \mu_B = T_{vv}- T_{uu} \, ,
\quad\quad
\partial_{r^*} \psi = 4 T_{uu}/(r-2M)\, .
\ee}
\be
\d s^2 = e^{\psi}\left[-\left(1 - {2 \tilde M \over r } \right) \d v^2 
+ 2 \d v \; \d r \right] + r^2 \d\Omega^2_2\, .
\label{twop}
\ee
where $\tilde M = M_0 + \mu(v,r)$ for $v > 0$. 

 In this new metric, the
matter action is the same as in equation  eq. (\ref{actS}):
\be
S^\phi_{g+h} = - \int  \! \d v\; \d r 
\left[\partial_v \phi \partial_r \phi 
+ \left(1 - {2 \tilde M \over r} \right) {(\partial_r \phi )^2  \over 2} 
\right] \, .
\label{actS2}
\ee
The new mass function $ \tilde M$ incorporates the only relevant metric 
change $\mu$.  Indeed, $S^\phi_{g+h}$ is independent of $\psi$, 
thereby recovering the 2D conformal invariance mentioned earlier.
 
Our aim is to work out the first order corrections due to the
gravitational interactions between $\phi_-$ and $\phi_+$.  To this end
only quadratic terms in $h$ should be kept in $S_{h, g}$.  The
Gaussian integration over $h$ can be performed (this is equivalent to
solve the linearized Einstein's equations). It gives rise to a
self-interacting field theory described by
\be
Z= 
 \int  \! {\cal{D}}\phi\, e^{iS^\phi_{g} + iS_{int}} \, .
\label{Z2}
\ee
By construction $S_{int}$ is a non-local quadratic form\footnote{In
the $t,r$ coordinate system, \ie, when $g_{rt}=0$, $S_{int}$ is given
by a linearized version (see equation  (90) in \cite{MP2}) of the
BCMN~\cite{bcmn} Hamiltonian.}  of the energy-momentum tensor of
$\phi$.
 
To identify the relevant part of $S_{int}$, 
we first recall that $T_{\mu \nu}$ has only two fluctuating components, 
thanks to the 2D conformal invariance.  
Thus, in a perturbative treatment (such as the interacting picture) 
one obtains two types of interaction terms only.
First one has self-interaction terms depending on $\phi_-$ or $\phi_+$
separately.  These terms do not destroy the factorisability of the
theory and will not be considered in what follows.\footnote{%
The validity of this 
simplification (also adopted
in \cite{THooft, Verl3, Casher}) requires further analysis. 
On-shell, the $\phi_+ \, \phi_+$ contribution to $S_{int}$
vanishes. This can be understood from the fact that the Vaidya metric
(2) is an exact solution for any classical infalling massless flux
$T_{vv}(v)$.
The $\phi_-\, \phi_-$ contribution to $S_{int}$ is more tricky to handle
in the advanced coordinates $v,r$.  The reason is that infalling
geodesics are affected by the presence of an outgoing flux $T_{uu}$
(as clearly seen when using the coordinates $u,r$). This modification
translates in $v,r$ into a deformation of the description of outgoing
geodesics $u=u(v,r)$ and it is this effect which is responsible for
the $\phi_- \phi_-$ contribution to $S_{int}$.  Let us finally notice
that a non-perturbative treatment of the self-interactions of $\phi_-$
has been developed in \cite{KKW}. It leads to small effects
$O(\kappa/M)$ and induces no damping of the waves when approaching the
horizon.}
Secondly, one has a cross term coupling $\phi_-$ to $\phi_+$.  
This term will inevitably break the
factorisability of the theory into the $\pm$ sectors. Therefore, 
the two-point function (\ref{8TT}) will no longer vanish.  
Let us analyze this term in more detail.

Since infalling configurations obey $\partial_r \phi_+ = 0$ even in
the presence of gravitational interactions, the cross term coupling
$\phi_-$ to $\phi_+$ is, see eq. (\ref{actS2}),
\ba
S_{int} &=& G
\int_0^{\infty} \!\!\d r \int_0^{\infty}  \!\! \d v  \; 
{ \mu_+(v) \over r } (\partial_r \phi_-)^2 \ , 
\nonumber \\
&=& G
\int_0^{\infty} \!\!\d r \int_0^{\infty}  \!\! \d v  \; 
{ \mu_+(v) \over r } \, ({du \over dr})^2 \, T_{uu} \, .
\label{sint2}
\ea
where $G$ is Newton's constant. 
We have re-introduced it in the front of $\mu$ to read more easily 
the order of the
interactions between $\phi_-$ and $\phi_+$ 
in the forthcoming equations.  $\mu_+(v)$ is the mass
fluctuation driven by $\phi_+$. Einstein's equations constrain is to be
\be
\mu_+(v) = \int_0^v  \! \! \d v' \; T_{vv}(v') 
= \int_0^v \d v' \; (\partial_{v'}\phi_+)^2 \, .
\label{mu1}
\ee
(The reader might be surprised by the fact that we are using 
on-shell fields $\phi_\pm$ in $S_{int}$, i.e. that we have used the equations
of motion. In principle
indeed, only the off-shell field $\phi$ should be used in the action.
However, when calculating perturbatively lowest order
corrections in $G$, this amounts to use eq. (\ref{sint2}) as it stands.)

We are now in position to show that the gravitational interactions
between $\phi_+$ and $\phi_-$ diverge on the horizon. To this end, let
us consider two classical fluxes described respectively by
$T_{vv} = \omega \, \delta(v - v_0)$ and $T_{uu} = \lambda \, 
\delta(u-u_0)$. $\omega$ and $\lambda$ are the asymptotic energies
measured on ${\cal J}^-$ and ${\cal J}^+$ respectively,
and $v_0$ and $u_0$ are such that the two spherical shells
meet at $r_0$ in the near horizon geometry, for $r_0 - 2M \ll 2M$.
In this case, using $r_0 - 2M
\simeq 2 M e^{\kappa(v_0 - u_0)}$, one obtains
\be
S_{int} \simeq 4G  { \omega \, \lambda \over r_0/2M - 1} \, .
\label{sintol}
\ee
The action $S_{int}$ diverges as $r_0 \to 2M$ like the FF frequency 
$\Omega$ did in (3).  
The difference with (3) is that $S_{int}$ is a {\it scalar}. Hence the
divergence in eq. (\ref{sintol}) is coordinate (gauge) invariant.

To get an estimate of where the gravitational interactions become strong,
\ie, can no longer be ignored, let us 
consider two shells whose asymptotic energy is Hawking temperature,
i.e. $\omega=\lambda= \kappa$.
The condition $S_{int} = 1$ is reached for 
\be
r_0/2M - 1 = 4G \,   \kappa^2\, .
\ee
The proper distance ($r_0 - 2M \simeq 1/M$) is much smaller
than Planck length, see footnote \ref{proper}.
This simple estimate will be recovered in Section 6 when considering 
radiative corrections in the {\it vacuum}.
We should 
emphasize this last point: even though our approach closely
follows that of \cite{THooft} (it can be considered as an $s$-wave
reduction of it) we shall not study the interactions between 
$\phi_+$ and $\phi_-$ quanta. Rather we shall focus on the {\it residual}
interactions when the state of $\phi_+$ is vacuum. For earlier
attempts in this direction, we refer to \cite{Verl3,Casher}.
Before analyzing these second quantized effects, it is instructive to 
solve two preparatory exercises with on-shell fluxes. 

In the first we shall show that 
$S_{int}$ acts only as a shift operator of the asymptotic value of $u$. 
In spite of this simplicity, in the second exercise, we show that 
$S_{int}$ nevertheless engenders an entanglement which prevents 
the factorisability of the states into $\pm$ sectors. This provides
an explicite exemple of a quantum effect induced by eq. (\ref{sint2})
which 
cannot be described in the semi-classical treatment.

\section{Non-vacuum gravitational effects}

\subsection{Classical interactions and shifts in $u$}

We first consider the following problem: 
Given two classical field configurations:
$\phi^0_+$ specified on ${\cal J}^-$ for $v> 0$ and $\phi^0_-$ on ${\cal J}^+$, 
what is the value of the field amplitude $\phi$ near the horizon
when taking into account the gravitational interactions of eq. (\ref{sint2}) ?

Because of the 2D conformal invariance, 
$\phi$ still decomposes as $\phi_+ + \phi_-$.
Then, since $\partial_r \phi_+ = 0$ is exact in our gauge 
wherein $v$ stays light-like, 
$\phi_+ (v)= \phi_+^0 (v)$ to all orders in $G$. 
Thus the infalling flux of energy is unaffected by the
energy carried by $\phi_-$ and it is given by its initial
value on  ${\cal J}^-$: $T_{vv}= (\partial_v
\phi^0_+)^2 $.  Hence, $\mu_+$ of eq. (\ref{mu1}) acts as a given metric
change in the equation of motion of $\phi_-$:
\be
\left[ 
2 \partial_v + \left(1 - { 2 M_0 + 2 G \mu_+(v) \over r }\right) \partial_r  
\right]
\phi_- = 0 \, .
\label{elag}
\ee
Since this equation is linear in $\phi_-$ and first order in the
space-time derivatives, its exact solution is
\be
\phi_-(v,r)= \phi_-^0(u_{\mu}(v,r))\, ,
\label{eik}
\ee
where $u_{\mu}(v,r)$ gives (in the coordinate system
of eq. (\ref{twop}))
the outgoing null geodesic in the modified
metric characterized by $M_0 + G \mu_+(v)$.

To determine the function $u_{\mu}(v,r)$, one notices that
it 
also obeys eq. (\ref{elag}) with
the (final) boundary condition that it converges to the un-modified value 
$u_{0}(v,r)= v - 2 r$ for $r \to \infty$.  
Hence, to first order in $G$, the change
$\delta u= u_{\mu} - u_0 $ is determined by a non-homogeneous
equation. 
Using the fact that $2 \partial_v + (1 - { 2 M_0/
r}) \partial_r $ defines $2\partial_v\vert_{u_0}$ (by definition of
$u_0(v,r)=constant$), $\delta u$ obeys\cite{Barr}
 \be
\partial_v\vert_{u_0}\delta u
= G \, { \mu_+ \over r} \, \partial_r\vert_v u_0 \, .\ee
The solution is
\ba
\delta u(v)\vert_{u_0} &=& G \int^\infty_{v}\!\! \d v' {2 \mu_+(v') \over 
r(v')\vert_{u_0} - 2M_0} \, ,
\label{17}
\ea
where $r(v)\vert_{u_0}$ is obtained by inverting $u_0(v,r)= v - 2
r^*$.  
As in the action (\ref{sint2}) the integral in
eq. (\ref{17}) is dominated by the near horizon region when the 
denominator $r - 2M_0 \ll 2M_0$. A tiny infalling energy flux
$\mu_+ \ll M$ can therefore induce an arbitrary large change of $u$.

The lesson we got from eq. (\ref{eik}) is that the eikonal approximation
is exact: the scattered value of the field amplitude is given by its
asymptotic value evaluated along the modified characteristic
$u_{\mu}(v,r)$.  Thus, classically, the gravitational interactions
encoded in eq. (\ref{sint2}) only induce a shift of the argument 
of field. They do {\it not} induce non trivial non-linearities in the
field amplitude in that $\mu_+(v)$ can be computed irrespectively 
of the value of $\phi_-$. The origin of this miracle is the 2D conformal
invariance. 

\subsection{Entanglement between ${\bf \phi_-}$ 
and ${\bf \phi_+}$} 

In spite of this absence of non-linearities in 
field amplitude, we shall now prove that 
the quantum evolution governed by the action
$S_g^\phi + S_{int}$ dynamically engenders entanglement between the
otherwise uncorrelated $\pm$ sectors.  To this end, we consider
the evolution 
of an initially factorized
wave function \be
\ket{\Psi^{in}}=\ket{\Psi_+^{in}} \otimes
\ket{\Psi_-^{in}}\, .
\ee  The infalling part $\ket{\Psi_+^{in}} $ is
specified on ${\cal J}^-$ for $v>0$. 
To clearly exhibit 
the entanglement, we choose $\ket{\Psi_+^{in}} $ to be a superposition 
of two well defined and separated states: 
\be
\ket{\Psi_+^{in}} = A \, \ket{\Psi_+^{in, \, a }}+ 
B \, \ket{\Psi_+^{in, \, b }} \, .
\label{supperp}
\ee
The two kets are normalized and orthognal to each other: $\langle\Psi_+^{in, \, i }
 \ket{\Psi_+^{in, \, j }}= \delta^{ij}$. Thus $A, B$ are probability 
 amplitudes obeying $\vert A\vert^2 + \vert B \vert^2 = 1 $.
By well defined and separated we mean that the two fluxes associated
with each component, $\bk{T_{vv}^i} \equiv \bra{\Psi_+^{in, \, i}}
T_{vv} \ket{\Psi_+^{in, \, i}}$ with $i=a,b$, are well localized
in $v$ and separated
from each other. 

The other 
piece of the initial ket, $\ket{\Psi_-^{in}}$, 
is specified on $v < 0$. After reflection on $r=0$,
it determines the state of outgoing configuarations. 
For the moment we do not need 
to further specify its quantum state. 

Having specified the initial state, we study the quantum dynamics
(in the interacting picture). 
The evolution operator $e^{i S_{int}}$ acting on the initial state
$\ket{\Psi^{in}}$ 
gives two {uncorrelated} evolutions weighted by $A$ and $B$:
\ba
\ket{\Psi} &\equiv& e^{i S_{int}} \ket{\Psi^{in}}
=   e^{i S_{int}} \Big[ \Big(
A \, \ket{\Psi_+^{in, \, a }}+ 
B \, \ket{\Psi_+^{in, \, b }} \Big) \otimes  \ket{\Psi_-^{in}}
\Big]\, ,
\nonumber\\
&=& A \,   e^{i S_{int}}\Big[\ket{\Psi_+^{in, \, a }} \otimes 
\ket{\Psi_-^{in}}  \Big]
+ B \,   e^{i S_{int}}\Big[ \ket{\Psi_+^{in, \, b }} \otimes
\ket{\Psi_-^{in}} \Big]\, .
\label{ent}
\ea
One should thus study each piece {\it separately}. This is nothing but the expression of the superposition principle.

In each state, the equation $\partial_r \phi_+ =0 $, now viewed as an
Heisenberg equation, tells us that the evolution in the $+$~sector is
trivial, as in classical terms. 
To study the evolution of the $-$~sector, 
we perform the approximation which consists in 
neglecting the fluctuations of $T_{vv}$ 
in each infalling state $\ket{\Psi_+^{in, \, i}}$. 
In this approximation, 
the evolution operator $e^{i S_{int}}$ becomes a $c$-number
for infalling configurations and acts only on 
outgoing configurations. One thus has
\ba
\ket{\Psi} &=& A \,  
\Big( 
\ket{\Psi_+^{in, \, a }} \otimes 
e^{i S_{int}^{a}} \ket{\Psi_-^{in}}\Big)
+ B \, \Big( \ket{\Psi_+^{in, \, b }} \otimes  e^{i S_{int}^{b}} 
\ket{\Psi_-^{in}}\Big)\, .
\label{ent2}
\ea
In the $A$-weighted state, the operator 
$e^{i S_{int}^a}$ governs the propagation of $\phi_-$ 
in the $a$-modified metric 
characterized by the mass function $M_a = M_0 + G \mu_+^a$ with 
\be
\mu_+^a(v) = \int^v_0 dv'\, \bra{\Psi_+^{in, \, a }}\hat T_{vv}(v')
\ket{\Psi_+^{in, \, a }} \, ,
\label{mua}
\ee 
whereas in the $B$ state, one finds
 the $b$-modified metric characterized by $\mu_+^b(v)$.   
The evolution in each case is thus governed by 
 the time-ordered product of ''its'' evolution operator
acting on same initial $\ket{\Psi_-^{in}}$. Explicitly, 
the two hamiltonians acting on $\ket{\Psi_-^{in}}$ 
are given by eq. (\ref{sint2}) 
with the corresponding the c-number metric changes $\mu_+^i$,
$i=a,b$.  
In the above treatment, 
we took
 into account  the first moment of $T_{vv}$, $\bk{T_{vv}}^i$, 
the mean value in each state. We indeed neglected the highers moments
which govern the fluctuations in each state. 
In doing so, we 
accounted for the fluctuations of $T_{vv}$
which are due to the fact that the infalling state
 is a superposition, but only those. This will be clarified below. 


The entanglement induced by $S_{int}$ acts, as usual, as a measurement.
For the interested reader we recommend the very 
instructive reading of Chapter IV. in \cite{Gott}.
Consider, as in that chapter, 
the Stern-Gerlach experiment wherein 
the center of mass motion of an electron in a magnetic field
is governed by its spin projection along that field.  
The mapping from that situation to the present one is as
follows.  The two kets representing the 
spin projections are here played by the two
infalling states $\ket{\Psi_+^{in, \, i }}$. The center of mass wave
function is played by the outgoing wave function $\ket{\Psi_-}$ 
and the interaction hamiltonian is $S_{int}$ of eq. (\ref{sint2}). 
 The analogy works quite well when the initial outgoing wave 
function $\ket{\Psi_-^{in}}$ 
also describes a localized flux. Then, its ``image'' on ${\cal J}^+$ would be {\it 
either} a spot at $u_0+\delta u_a$ with probability $\vert A \vert^2$, 
{\it or} one at $u_0+\delta u_b$ with probability $1-\vert A \vert^2$.  
The location $u_0$ is that of the spot 
when the gravitational interactions are ignored, whereas the values of the
shifts $\delta u_a, \delta u_b$ are given by eq. (\ref{17}), 
evaluated along the unperturbed 
geodesic $u_0$, fed with the
mass changes $\mu_+^a$ and $\mu_+^b$ respectively, and with
the lower value $v$ set to $0$. 
 
This quantum result should be compared with what would have been
obtained by applying the semi-classical treatment to the entire
wave function (rather than to each sector separately).
In that case, the change in the common location is driven the {\it
mean} mass change 
\be
\bar \mu_+(v) = \vert A \vert^2 \, \mu_+^a(v) + \vert B \vert^2 \, \mu_+^b(v) \, .
\label{mumean}
\ee
Therefore, the semi-classical treatment {\it incorrectly predicts} a single spot on
${\cal J}^+$ located at the ``mean'' position $u_0+ \bar{\delta u}$
with $\bar{\delta u}= \vert A \vert^2 \delta u_a + \vert B \vert^2
\delta u_b$.  

In the above exemple, 
the validity of the semi-classical treatment rests on
the possibility of neglecting the fluctuations of the 
$\phi_+$-operator-valued shift $\hat{\delta u}$. The dispersion
about the mean quantifies their importance. One finds
\ba
\bk{\hat {\delta u}_{u_0} \, \hat {\delta u}_{u_0}}_C
&\equiv&\bk{\Big(\hat {\delta u}_{u_0} - \bk{\hat {\delta u}_{u_0}}
\Big)^2}
=  \bk{(\hat {\delta u}_{u_0})^2} - \bk{\hat {\delta u}_{u_0}}^2 \, ,
\nonumber\\
&=&
\vert A  B \vert^2 \, \Big(\delta u_a(u_0)
-\delta u_a(u_0) \Big)^2\, ,
\label{neweq}
\\
&=& 4 G^2 \int^\infty_{0}\!\! \int^\infty_{0}\!\! \d v_1 dv_2
\, {\bk{\mu_+(v_1)\mu_+(v_2)}_C \over 
(r(v_1)\vert_{u_0} - 2M_0)(r(v_2)\vert_{u_0} - 2M_0)} \, .
\label{17b}
\ea
To get the second line we have used the results 
of the above two paragraphs. From eq. (\ref{neweq})
one verifies that the semi-classical treatment
is valid iff 
\be
{\bk{\hat {\delta u}_{u_0} \, \hat {\delta u}_{u_0}}_C
\over \bk{\hat {\delta u}_{u_0}}^2 } = \vert A  B \vert^2 \, 
{ \Big(\delta u_a
-\delta u_a \Big)^2 
\over \Big(\bar{\delta u} \Big)^2} \ll 1\, .
\ee
It is negligeable either when $\vert A  B \vert^2 \to 0 $, that is,
when one of the infalling flux is rarely found, or 
when the two shifts in $u$ are close to their mean.\footnote{We refer to 
\cite{Parentani:1997ww,Massar:1997en,Massar:1998cs}
for a treatment of the solutions of the Wheleer-deWitt equation in which
the gravitational back-reaction is computed for each component of the matter
wave function separately. This radically differs from the usual mean 
field treatment in which the gravitational back-reaction is determined
at once\cite{Brout:1988ku}. 
Only the first treatment respects the supersposition principle. Indeed the
back-reaction in the second case contains the probabilities to find each 
component, exactly like 
in $\bar{\delta u}= \vert A \vert^2 \delta u_a + \vert B \vert^2
\delta u_b$. Hence 
the limitations of the validity of the second (the semi-classical treatment) 
are not intrinsic to the gravitational dynamics.} 

To get eq. (\ref{17b}) we simply have used the definition of $\delta u$,
eq. (\ref{17}), and that of 
connected correlation functions, see (\ref{8TT}).
Eq. (\ref{17b}) displays the relation
bewteen the dispersion of $\delta u$ and $\bk{\mu_+(v)\mu_+(v')}_C$,
the correlation function of the metric fluctuation
$\mu(v)$ in the quantum state one is dealing with,
here given in eq. (\ref{supperp}).\footnote{We remind the reader that to get
(\ref{neweq}) we have neglected the fluctuations of $\mu$ in each infalling state,
i.e. only the dispersion induced by the $A$-$B$ superposition has so far been
taken into account. The aim of the next Section is to incorporate
 the inherent fluctuations which are present in any infalling state including the 
 vacuum. 
  This explains the title of the
present Section: {\it Non-vacuum gravitational effects}.} 
  Because of the horizon,
the importance of the dispersion is determined by two things.
On the one hand, the fluctuations of $\hat{\delta u}$ are 
driven by the fluctuations of the 
mass function $\hat \mu = \int dv \hat T_{vv}$ which are themselves
determined by those of the infalling flux $\hat T_{vv}$, given the state
of infalling configurations. The  connected 
correlation function  $\bk{ T_{vv} T_{vv}}_C$ quantifies 
their dispersion.
On the other hand, the mass fluctuations are amplified by 
the denominators in eq. (\ref{17b}). 
Therefore, 
even if the relative importance of 
$\bk{T_{vv} T_{vv}}_C$
with respect to the square mean  $\bk{T_{vv}}^2$
is small, the semi-classical treatment can give incorrect
predictions
when the fluctuations are sufficiently
amplified by the denominators. 
Instead, in the absence of amplification, the mean theory provides a 
reliable approximation, unless of course if
$\mu_a(v) - \mu_b(v))/M \simeq 1$, but in this case 
the $a$,$b$ branches completely decouple/decohere,
and one should consider each possibility separately. 

In brief, the crucial points we have reached are the following.
First, unlike what one encounters in usual cicumstances, \ie, 
without an event horizon, 
tiny fluctuations of $T_{vv}$ might give rise to large
shifts in $u$ because of the amplification due to the growth 
of the gravitational interactions when the configurations
meet near an event horizon. Second, these fluctuations
break the factorizability of the theory into the $\phi_+, \phi_-$
sectors. Third, the quantity which governs the validity of the
semi-classical scenario is $\bk{T_{vv} T_{vv}}_C$,
the connected two point function of  $T_{vv}$.

The aim of Section 5 is to extend the analysis of
eq. (\ref{17b})
when the infalling configurations are in the vacuum, i.e. in
 a coherent state as opposed to a superposition as in (\ref{supperp}).
In that Section, the (inherent, minimal, and divergent)
vacuum fluctuations will be the only contribution to
$\bk{T_{vv} T_{vv}}_C$.

\vskip .3 truecm

\subsection{Relationship with former treatments}

\noindent
Before considering vacuum effects, it is also interesting to
relate eq. (\ref{Z2}) to the former treatments of black hole evaporation
discussed in the literature: 
Hawking's approach~\cite{Hawk} and the semi-classical treatment.

Hawking's approach formulated in a fixed geometry
is recovered by putting $G\mu_+ =0$ in
eq. (\ref{Z2}). Then $Z$ factorizes as $Z_+\otimes Z_-$ 
(when ignoring the trace anomaly) 
and $\phi_-$ is a free outgoing field propagating in
the background geometry $g$. Thus $\phi_+$ drops out from {\it
all} matrix elements built with the operator $\phi_-$.  It should be
emphasized that the trans-Planckian problem 
encountered in Hawking's approach directly
follows from this factorisability. Indeed it is the {\it absence} of
gravitational coupling between the $+$ and $-$ sectors which permits
the unbounded growth of frequencies when probing, near the horizon,
configurations specified on ${\cal J}^+$\cite{MP2,Barr}.

The semi-classical treatment\cite{Bardeen,PP,massar} 
can be obtained from the path integral
formalism eq. (\ref{Z1}) by first integrating over $\phi$ at fixed $h$ and
then searching for the classical extremum of $h$.
In this approach, by construction, the fluctuations of $h$ and 
$T_{\mu \nu}$ are neglected. Thus the near
horizon propagation of $\phi$ is governed by a single but now
{\it self-consistent} metric governed by mean $ \langle \mu_+(v) \rangle$.  
This mean
evolution characterizes by the shrinking of the horizon area 
according to 
\be
{ d\langle \mu_+(v) \rangle \over dv } 
= \langle T_{vv}\rangle \vert_{r=r_{horizon}=2M} \,\, .
\label{one}
\ee
When working in the vacuum eq. (\ref{phi2p}), the 
(properly
subtracted~\cite{GO}) expectation of $T_{vv}$ is
\be
\langle T_{vv}(v) \rangle\vert_{r=2M(v)} =
 - {  \pi  \over 12 } \left({\kappa(v) \over 2\pi}\right)^2  \, ,
\label{meanT}
\ee
where $\kappa(v)= 1/4 M(v)$ with $M(v)= M_0 + G \bk{\mu_+(v)}$.
This flux has the opposite value of a 2D thermal flux.
The only change with respect to the fixed background
approach of Hawking is the replacement of $M_0$ by $M(v)$.
Thus the propagation of out-going configurations is
hardly affected by the evaporation as long as it is slow, \ie, as long
as $ M(v) \gg M_{Planck} $.

Therefore, in the semi-classical scenario, the trans-Planckian problem
stays as in Hawking's approach: The coupling between $\phi_-$ and the
mean change $\langle \mu_+ \rangle$ is incapable to provide a taming
mechanism since it does not open new interacting channels.  To solve
this problem clearly requires to take into account the {\it
fluctuating} character of the interactions between $\phi_-$ and
$\phi_+$, \ie, the possibility of entangling their wave functions, as
in the quantum mechanical exercise presented above.

\section{Modified two-point function}

Our aim is to determine how the (dressed) two-point 
function of $\phi_-$
\be
G_-(x_2; x_1) = { \int {\cal D} \phi \; \phi_-(x_2)\; \phi_-(x_1) \;
 e^{iS_g+iS_{int}}   \over Z} \, ,
\label{Gmod}
\ee
where $Z$ is given in eq. (\ref{Z2}), is affected by the gravitational
interactions encoded in $S_{int}$ 
when the infalling configurations are in their vacuum
state.

To evaluate eq. (\ref{Gmod}) beyond the semi-classical
treatment, one should adopt some rules to cope with the UV divergences. 
Even at one loop, there are several inequivalent graphs
which result from the various Wick contractions. To extract
consistently the contribution of the simplest ones, 
we propose to 
consider $N$ copies of $\phi$.
The calculation of eq. (\ref{Gmod}) 
can then be achieved in two different (but equivalent) approaches. 
The first consists  integrating first over the $N-1
\simeq N$ spectator copies not appearing in the numerator in
so as to determine the (linear term in $N$ of the) influence functional (IF) \cite{FHibbs} governing the effective dynamics of $\phi$.
The other consists in
developing $e^{iS_{int}}$ in both integrands of eq. (\ref{Gmod}) in powers
of $S_{int}$ so as to engender the (connected) graphs
governing the radiative corrections. Then in the large-$N$ limit the
graphs can be classified into infinite series according to the relative power
between $G$ and $N$, see the Appendix for the details. The leading
series (in $GN$) reproduces the semi-classical treatement. The next series 
of graphs, those weighted by powers of $G^2 N$, all correspond to the
graphs engendered by the term in the influence functional
which is linear in $G^2 N$.

The usefulness of the IF approach is that non-linear (in $G^2 N$) modifications 
of the propagation of outgoing configurations  are
easily taken into account through this influence functional.
The same results 
can of course be
reached from the diagrams approach at the cost of re-summing
the corresponding infinite subsets of graphs. 
It is in the {\it identification} of these
infinite subsets that the large $N$ limit finds its 
justification. For a schematic description of these diagrammatic
aspects, we refer to the Appendix.  
In what follows, we shall pursue with the IF approach.

When computing the lowest order corrections to the effective action in
eq. (\ref{Gmod}), we can use eq. (\ref{phi2p}),
the `free' propagator of $\phi_+$.\footnote{
In would also be interesting to determine if
the higher order terms in 
powers of $G$ 
will become operator-valued~\cite{Verl3} in $\phi_-$, thereby obtaining a
situation analogous to that of transition amplitudes when enlarging
the phase space so as to take into account recoil effects~\cite{rec}.} 
This approximation concerning degrees of freedom {\it not} 
directly involved in the matrix elements
(\ie, which factorized out in the absence of interactions) is a common
procedure both in quantum field theory where it gives the vacuum
contribution, see chapter 9 in \cite{FHibbs}, and in statistical
mechanics (\eg, the {\it polaron}, chapter 11).  In our case, in this
approximation, the IF gives rise to a non-local action which is a sum
of terms containing $(\partial_r \phi_-)^2$ and kernels given by the
Wick contractions of $T_{vv}$ evaluated with eq. (\ref{phi2p}).
To order $G^2 N$, one obtains\footnote{\label{weinbrem}\it{In the above expression,
we did not take into account the
time ordered character of the evolution operator 
$e^{i S_{int}}$. It would be interesting to determine the 
consequences of this neglect. We refer to App. A of \cite{MaPa1}
for a discussion of this point in the context of atomic transitions.
We also refer to the recent work of Weinberg\cite{Weinb} wherein
 the 1-loop radiative corrections to 
the two-point function determining the power spectrum in a 
inflationary context are computed (in a large $N$ limit as well) taking properly
into account the time ordering. 
In \cite{Parentani:2002mb}, we tried to improve the calculation of 
\cite{hawkm} by taking it into account. We obtained an extra divergence 
in eq. (\ref{res}) which induces dispersive effects. However the meaning of this
extra term is unclear 
as it seems to correspond to a tadpole contribution which should be subtracted.}}
\be
S_{IF} 
= i \langle S_{int} S_{int} \rangle_+ = i G^2N\, \int \!\! d^2 x\!
 \int \!\!d^2x'
 (r r')^{-1} (\partial_r \phi_-)^2 \, \langle \mu_+(v) \mu_+(v') \rangle\,  
(\partial_{r'} \phi_-)^2 \, ,
\label{effact}
\ee
where $\langle \quad \rangle_+$ means that the expectation value
applies to $\phi_+$ only.
Using eq. (\ref{phi2p}), the connected two-point function is
\be
\langle  T_{vv}(v)  \; T_{vv}(v') \rangle_C = {1 \over 16 \pi^2 }
{1 \over (v-v'- i \epsilon)^4 } \, .
\label{Tfluct}
\ee
Then, eq. (\ref{mu1}) gives
\ba
\langle  \mu_{+} (v)  \; \mu_{+} (v') \rangle &=& {1 \over 96 \pi^2 } 
{1 \over (v-v'- i \epsilon)^2 } 
\nonumber\\
&=& {1 \over 96 \pi^2 } \int_0^\infty \!\! \d\om \; \om  
\exp[-i \om(v-v')] \, .
\label{mf2}
\ea
This equation gives the mean 
fluctuations driven by one $\phi_+$ field in
the unperturbed ($G=0$) vacuum state, see \cite{Verda} for a analysis
of the 4D two-point function of induced metric fluctuations in
Minkowski vacuum, see also \cite{HuR} for a analysis of metric fluctuations
in the near horizon region.  

It should be noticed that $\langle \mu_{+} (v) \mu_{+} (v')
\rangle $ is not real. This follows from the quantum vacuum which
contains only positive frequencies when hit by $\phi$, see
eq. (\ref{phi2p}).
Notice that eq. (\ref{Tfluct}) gives the ``un-subtracted'' value of the 
connected two-point function. As shown in \cite{Tomb}, 
the counter-terms which lead to the renormalized one-point 
function, eq. (\ref{meanT}), provide divergent contributions
to eq. (\ref{Tfluct}) which tame its singular behaviour as $v \to v'$,
and make it a well-defined distribution\cite{Verda}.

Keeping only eq. (\ref{effact}) 
in the IF (or by summing the corresponding infinite set of Feynman graphs, 
see the Appendix) is equivalent to
work with a stochastic (\ie, a classically given, albeit not real) and
Gaussian ensemble of metric fluctuations.  
By equivalent we mean that the graphs obtained from the Gaussian
integration in the stochastic treatment
are in one to one correspondance of those of the $G^2 N$-truncated 
quantum treatment. Therefore all matrix elements
of $\phi_-$, such as eq. (\ref{Gmod}), can be computed from the
stochastic theory. 
In brief, as far as the 
propagation of $\phi_-$ is concerned, neglecting the time ordering
in the evolution operator, and to lowest order in $G$, 
the functional integration over $\phi_+$ in eq. (\ref{Gmod}) 
defines an effective stochastic 
ensemble of metric fluctuations governed by eq. (\ref{mf2}).

In this case, all the techniques developed in~\cite{Barr}
apply. In what follows we shall present schematically the main
results and we refer to this work for details.
The key point is the following.
Because of the Gaussianity of the ensemble, 
one can obtain {\it non-linear} corrections to eq. (\ref{Gmod}) from the 
fluctuating characteristics of eq. (\ref{elag}), \ie, the outgoing null 
geodesics $u_\mu(v,r)$, the non-trivial 
solutions of $\d s^2=0$ in the fluctuating metric eq. (\ref{twop}).
In this we recovered that there is no non-linearities in the 
field amplitude: as in Section 4 the non-linearities in $G$ only occur 
through the characteristics. 

To determine the effects engendered these metric 
fluctuations, it is instructive to analyze the backward in time 
propagation of configurations representing asymptotic Hawking quanta. 
In particular it will reveal the role played by 
shift in $u$ studied in section 4.1.
To this end, we consider
 the Fourier transform, performed on ${\cal J}^+$, of the $in-in$ Green function
(that obtained by taking the expectation value in the initial vacuum
on ${\cal J}^-$): 
\be
G_-(\lambda; v,r) \equiv \int du_1 \, e^{i \la u_1} \, G_-(v_1=\infty,u_1; v,r) \, .
 \label{FT}
\ee 
In the unperturbed metric, $\mu=0$, i.e. ignoring $S_{int}$ in eq. (\ref{Gmod}),
we get
 \be
G_-(\lambda; v,r) \propto e^{i \la u(v,r)} \, ,
\ee 
where
 $u(v,r)=u_0(v,r)= v-2r^*$ is the unperturbed null characteristic.
Hence near the horizon, the Fourier transform of $G_-$
behaves as
\be
e^{-i \la u_0(v,r)} \simeq 
e^{-i \la v} \; (r-2M_0)^{i\kappa \la} \, .
\label{19}
\ee
It 
possesses an infinite number of
oscillations as $r \to 2M_0$ with a radial 
momentum given by \be
\hat p_r \, e^{-i \la u_0(v,r)} = -i
\partial_r \, e^{-i \la u_0(v,r)}=  {2\la \over r/2M_0 - 1 }\,\,
 e^{-i \la u_0(v,r)} \equiv p_0(v,r) \, e^{-i \la u_0(v,r)}\,  .
\label{pr}
\ee 
A FF observer will attribute to this wave a FF frequency that
grows as in eq. (\ref{dopef}) because
 $dr\vert_v \propto d\tau$, where $\tau$ is his proper time.
 In fact $p_r$ should be interpreted as a frequency rather than 
 a momentum (its sign fixes that of the Klein-Gordon inner product
 \cite{bmps,Balbinot:2006ua}).
 

Taking into account the first order change in
$u$, see eq. (\ref{17}),
in any Gaussian ensemble of infalling metric fluctuations, the 
Fourier transform of the ensemble average of the $in-in$ Green function of
eq. (\ref{FT}) %
will be governed by the averaged waves given by
\be
\langle\langle \, e^{-i \la u_\mu(v,r)} 
\,  \rangle \rangle = e^{-i \la u_0(v,r)} \; 
e^{- {\la^2 \over 2}\langle\langle \delta u(v)\delta u(v) \rangle\rangle  } \, .
\label{18}
\ee
In the above the averaging acts on the fluctuating
quantity $\mu_+(v)$.

We now focus on the particular (Gaussian) ensemble wherein the 
fluctuations are those induced by $N$ fields in the vacuum, 
i.e. they are governed by $N$ times the expression of eq. (\ref{mf2}).
Then, using the fact that $r_0(v)\vert_{u_0} -2M_0 
\simeq 2M_0 \, e^{\kappa(v-u_0)}$, one obtains 
\ba
 \langle \delta u(v)\vert_{u_0} \; \delta u(v)\vert_{u_0} \rangle &=& 
{ G^2N \over (r /2M_0 -1)^2} \int_0^\Lambda {\d\om
\over 3} 
{\kappa^2 \om \over \kappa^2 + \om ^2 }
\nonumber\\
&=&
 \bar \sigma_\Lambda^2 \,  {1 \over (r /2M_0 -1)^2}
\, . \quad \quad
\label{res}
\ea
The spread $\bar \sigma_\Lambda$ governs the damping of the backward
propagated waves.  It is equal to 
\be
\bar \sigma_\Lambda = 
G\kappa \sqrt{N\,  \ln(\Lambda /
\kappa)/3 }\, , 
\label{bars}
 \ee
 when the hard UV cut-off $\Lambda$ satisfies $\Lambda \gg
\kappa$.  We have introduced $\Lambda$ to define the two integrals over
$\om$. Notice that $\Lambda$ is a Lorentz scalar 
in the sense that it is 
the energy of an $s$-wave in its rest frame in a stationary and spherically
symmetric background. 

The important result of eq. (\ref{res}) is that $\bar \sigma_\Lambda$ is not
proportional to $\Lambda$ even though $\langle \mu_+^2 \rangle \simeq
\Lambda^2$.  Notice indeed that $\bar \sigma_\Lambda$ hardly depends on
the value of $\Lambda$ since \be
\bar \sigma_{\Lambda=4M_{0}} = \sqrt{2} \
\bar \sigma_{\Lambda=M_{Planck}}\, .
\ee  This insensitivity follows from the fact that high frequencies ($\om \gg
\kappa$) are damped by the integration over $v'$ in eq. 
(\ref{17}).  The
frequencies $\om \simeq \kappa$ dominate the contribution to $\bar \sigma$ in
eq. (\ref{res}), see \cite{Weinb} for a similar result in an inflationary 
context.
However they are not sufficiently damped to give a finite result.
Therefore, in the simplified treatment we are using, the
value of $\Lambda$ must be chosen from the 
outset.\footnote{\it In spite of this ambiguity, we still believe that 
the $\om \simeq \kappa$ dominance and the associate robustness of 
$\bar \sigma$ give credit to the validity of (\ref{bars}).
This is conforted by the fact that $\bar \sigma$ also
governs the modifications of the asymptotic properties of 
Hawking radiation\cite{Barr}.)}  Instead, in an
improved treatment 
of the regularization of the divergences in eq. (\ref{res}),
we believe that this ambiguity will be resolved.
That is, 
when using
the properly subtracted \cite{Tomb} two-point function $\bk{T_{\mu \nu}
T_{\alpha \beta}}$,  
only $\om \simeq \kappa$ should contribute to the $\bar
\sigma$'s. The reason is the following:
in the UV domain, all expressions become Minkowskian
in character and hence cannot contribute to Lorentz breaking effects
such as those engendered by $\bar \sigma$. 

The second result of eq. (\ref{res}) is that $\langle \delta u\; \delta u
\rangle $ diverges as $r \to 2M_0$. Thus the correlations between
asymptotic quanta and early configurations, which existed in a given
background as shown in eq. (\ref{19}), are washed out by the metric
fluctuations once $r - 2M_0 \simeq \bar \sigma_{\Lambda}$.
The reason of this loss of coherence is that the state of
$\phi_-$ becomes correlated to that of $\phi_+$~\cite{THooft,Verl3}.
Physically, this loss of coherence implies that induced emission
\cite{wald} no longer exists when the threshold energy (measured in the FF frame)
$1/\bar \sigma= 4 M_0$ is reached.\footnote{\label{lost}An interesting and unsolved 
question raised by this loss is whether {\bf new} correlations are 
induced by the gravitational interactions as the same time as the 
old ones are washed out. 
{\it This phenomenon of replacement of correlations has been
clearly derived 
in a slightly different context in \cite{ModeC}}.}
Phenomenologically this loss can be viewed as a dissipation of
outgoing waves, and, as in condensed matter~\cite{rmed,hu},
it can be described by an {\it effective} 
dispersion relation. 
Explicitely, using the notations of \cite{Balbinot:2006ua,Jacobson:2007jx}, 
 we get an effective relation $
 \Omega^2 = F^2(p)$,
 where $\Omega$ and $p$ are respectively the frequency and the radial
 momentum measured in the FF frame, and where $F^2$ behaves as
 \be
 F^2(p)= p^2 \, ( 1 - 2 i \, {\bar \sigma^2 \kappa p }) =  p^2 \, ( 1 - 2 i \, 
 {p \over p_c} )\, .
 \ee
 To obtain this, it suffices to rewrite the r.h.s. of (\ref{18}) as
$ e^{-i \la u_0} \, e^{-\bar \sigma^2  p_0^2/2}$ and to work in the momentum representation. Then with the help of eqs. (123, 124) in \cite{Balbinot:2006ua} the
  identification of the function $F$ is straightforward. It should be 
  noticed that the $UV$ momentum scale $p_c$ which weighs the cubic term is 
  $p_c = (\bar \sigma^2 \kappa)^{-1} = (1/\bar \sigma) \, 
  (L_{Pl} \kappa)^{-2}  = M_{Pl} \, (L_{Pl} \kappa)^{-3}$, and {\bf not}
  $p_c =1/\bar \sigma$ as one might have thought.
  This mismatch of $UV$ scales illustrates that it is probably meaningless 
  to try to identify 
  a well defined dispersion relation when
  starting from the result: the modified propagation of eq. (\ref{18}) which 
  unequivocally states that backward propagated modes are dissipated when
  their FF momentum reaches $1/\bar \sigma$. The reason why this 
  identification fails is that the modified propagation 
arises from the non-trivial properties of the background and not 
  from short-distance physics as it is the case in condensed matter.

We should further explain the physical relevance of these results. 
To this end, one must identify the matrix elements of
$\phi_-$ which are 
sensitive to the metric fluctuations (and
governed by the ensemble averaged waves eq. (\ref{18})) and those which
are not.  The simplest exemple of an operator which is sensitive is
the Fourier transform of the $in$-$in$ two-point function 
in eq. (\ref{FT}). 
On one hand, the phase of the wave function used in the Fourier transform
is not affected by the metric fluctuations.
On the other hand, the location $v,r$ of the second field operator 
is also unaffected by the metric fluctuations. However the two-point 
correlation function
 is sensitive to the metric fluctuations
encountered from ${\cal J}^+$ to $v,r$.\footnote{One might 
wonder if the effects we are describing are not induced
by the choice of working at fixed $u$ or at fixed $v,r$. 
To avoid misunderstanding, we recall that Green functions have no physical meaning
{\emph{per se}}, rather they are elements which appear in integrals
describing transition amplitudes (for a discussion of this point in a
quantum gravitational context see Section 2 in \cite{wdwgf}). 
Having made this remark, one verifies that $u$ is a physically
meaningful coordinate on ${\cal J}^+$ by noticing that $\d u\vert_r = \d t$ where
$\d t$ is the proper time of a particle detector on ${\cal J}^+$.
Thus when additional quantum mechanical systems are coupled to the
radiation field, the matrix elements governing transition amplitudes
will have, in their integrand, phase factors behaving like
$e^{-i\lambda u}$ in {\it any} coordinate system.  Similarly, upon
questioning what an infalling observer might see when crossing the
horizon, the dependence in $v,r$ is meaningful 
since $\d r\vert_v \propto - \d\tau$
where $\d\tau$ is his proper time.}  It
is this ({\it unusual}, see below) discrepancy between 
the phase at each point which explains why the ensemble averaged
 waves of eq. (\ref{18}) govern the above Fourier transform
of the two-point function.

Instead {\it usual} expectation values, such as
for instance the $in$--$in$ Green function with two points evaluated
at fixed $u$ on ${\cal J}^+$ (or two nearby points close to the
horizon), are {\bf not} severely affected by the metric fluctuations because
these expectation values
are not governed by the ensemble averaged waves (\ref{18}).  The
reason is that the ensemble average is performed {\it after} having
computed the $\phi_-$ expectation value 
for each member of the ensemble. (This is not a
choice: the stochastic ensemble is merely a tool to
{reproduce} quantum mechanical expectation values.  Its quantum
origin fixes the rules of ensemble averaging without
ambiguity.). In our case, this implies that the shift eq. (\ref{17})
affects {\it coherently} the phase at each point, see Section IV.A in
\cite{Barr}.  This guarantees that the shift drops out in the
coincidence point limit.  This cancellation in turn guarantees that
the asymptotic properties are (almost) unaffected since the Green
function possesses the usual Hadamard singularity~\cite{new}. By
almost we mean that the corrections scale like $(\kappa \bar
\sigma)^2$ and thus are order $1/M^4$. It is important to 
point out that
it is again the dynamically induced scale $\bar \sigma_\Lambda$ and not the
UV cut-off $\Lambda$ which governs these corrections.

We would like to further discuss the fact that the metric
fluctuations strongly affect the correlations between configurations
specified on ${\cal J}^+$ and near the horizon without modifying the
short distance behaviour of the Green function.\footnote{This clearly 
illustrates that the physics seen by infalling observers completely differs 
from that reconstructed from observers at large distance from the hole. 
This is similar to what 
was advocated in~\cite{Verl3}. However, 
the fact that the FT transformed on ${\cal J}^+$  is governed 
by eq. (\ref{18}) indicates that the near horizon physics
is unaccessible and therefore lost to remote observers.
Thus it seems that these two descriptions cannot not
obey the `complementarity principle'~\cite{Verl3}. 
Indeed,  it appears from our analysis that in the interacting FF vacuum
early out going configurations are entangled to infalling ones.  
Let us recall that 
by complementary to each other, it was meant that the 
two descriptions are both complete, like the position
and momentum representations of the same vector state in quantum
mechanics.}
 The radical 
difference of the impact of vacuum gravitational interactions follows
from the fact that any asymptotic measurement
will 
 always 
 be governed by projectors on some $out$-states, 
 i.e. states with a definite particle content defined
 on ${\cal J}^+$, to
probe the physics. Therefore, these measurements 
will always be of the $in-out$ type since the Heisenberg state of the
field is specified (prepared) before the collapse. It is this
two-states formalism giving rise to non-diagonal matrix
elements~\cite{MP2} (exactly like in a $S$-matrix
formulation~\cite{THooft}) which is at the origin of the
difference: The metric fluctuations cannot coherently affect
configurations specified in the `ket' on ${\cal J}^-$ and in the `bra'
on ${\cal J}^+$, hence the coherence is lost.
On the contrary, measurements performed by 
infalling observers only probe 
the near horizon behaviour of the Green function. 
Hence the coherence is maintained for them.

\section{Conclusions}

We have studied the effects induced by the gravitational interactions
governed by eq. (\ref{sint2}).  Even though we worked out only the lowest
order in $G$ ($\bar \sigma_\Lambda \propto G$) we believe that our 
main result is robust:\footnote{\it
{\underline{Added comment}}: 
We can not exclude the possibility 
that, when properly renormalized, the 1-loop radiative corrections governed by
(\ref{effact}) will give $\bar \sigma_\Lambda =0$. In fact, the self-energy
of an electron in a thermal bath of 
photons 
(as a thermal bath of gravitons) induces no
 dissipative effects at one loop, see \cite{APV} and refs. therein.
Nevertheless the naive
 reasoning that loop corrections in a thermal bath induce dissipative
 effects 'becomes' correct at two-loop level. These effects 
 are indeed propotional to $e^4$ where $e$ is the electric charge.
Thus, in the case one finds $\bar \sigma_\Lambda =0$ at one loop, one 
still needs to confront the question whether this result
be found at any loop level, to all order in $G$ (for a deep reason, e.g. 
because we are dealing with the 
true interacting vacuum), or shall $\bar \sigma_\Lambda \neq 0$ at some
higher loop level
(as in a heat bath, because the integrands are no longer 2D Lorentz invariant), 
thereby validating the present conclusions.}  We see no reason 
for higher order terms to {\it suppress} the entanglement of $\phi_-$
with $\phi_+$ so as to give $\bar \sigma_\Lambda =0$ thereby {\it
recovering} trans-Planckian correlations.  Indeed higher order
modifications to eq. (\ref{Tfluct}) and eq. (\ref{mf2}) should be of the type
$(G\omega^2)^n$ and therefore will not affect
the low frequency behaviour of eq. (\ref{mf2}) thereby leaving the
effective spread $\bar \sigma_\Lambda$ essentially untouched.
Moreover, when considering the effects of higher angular momentum
modes, as indicating by \cite{Casher}, $\bar \sigma$ should be {\it
larger} than our estimate based on s-modes because
the effects of higher angular momenta should add incoherently.

In brief, given that gravitational interactions grow without bound near the
horizon, we claim that the entanglement of $\phi_-$ with $\phi_+$ 
is unavoidable and universal. (By universal
we mean that a similar entanglement would be also 
found when considering the coupling of $\phi$ to other 
quantum fields such as massive ones.)  
The entanglement will then prevent the unbounded growth of 
frequencies encountered in the free field
theory and will be accompanied by the reorganization of the
description of vacuum.  That is, when $r \to 2M$,
the usual states of the free field theory which give rise 
to the notion of on-shell asymptotic
particles provide bad approximations of the true eigenstates
(albeit still characterized by the Killing energy $\lambda$ 
since the situation is stationary)
of the interacting theory. 
It is the growing discrepancy which leads to the `dissipation' 
of the amplitude in eq. (\ref{18}).\footnote{\it Motivated by the present
work, we have recently constructed an interacting QFT model
which is exactly solvable and which explicitly displays this 
phenomenon\cite{ModeC}, namely the growing discrepancy between the usual
modes and the true eigenstates manisfests itself through
the dissipation of the two-point function.}

We also claim that the radiative corrections encoding dissipation
should be finite because only low frequency
$\omega \simeq \kappa$ infalling configurations contribute to them. 
Indeed, in the UV regime for both infalling and outgoing configurations,
the expressions coincide with those evaluated in the tangent plane 
and are Minkowskian in character. 
Hence the high frequency regime cannot contribute to the effects
which break Lorentz invariance. 
(This still needs to be confirmed by an explicit
calculation.)

We would like to conclude this work by several remarks on related
aspects of Quantum Gravity and Black Hole Physics.

First, we point out the similarity between the above effects induced by 
the metric fluctuations representing gravitational interactions in the
vacuum and those attributed~\cite{garay} to `foam'.  By foam we mean
quantum gravitational configurations which radically affect the
smoothness of space-time at short distance. (They might arise from
gravitational instantons~\cite{hawkk} or stringy
effects~\cite{kempfcit}.)  In all cases, the replacement of free field
propagation in a fixed background by the appropriate interacting model
might lead to very similar (universal?) deviations when analyzing the
departure from the free field description that all models possess at
large distances. Therefore, these first deviations might be described by 
some effective mesoscopic theory of space-time properties which would
essentially signal the existence of a minimal resolution 
length~\cite{kempf}, the equivalent of our $\bar \sigma$,  
in the otherwise local field theory.

Secondly, we conjecture that $\bar \sigma$ 
(properly computed so as to include the contribution of
higher angular momentum modes) should
{\it also} be the length scale which enters in the entanglement
description of the black hole entropy~\cite{bhee}.  We recall that when
using free field in a given space time, the entanglement entropy 
diverges due
to the unbounded character of the reservoir of high energy modes.  To
get a finite entropy density per unit area, some cut-off should be
introduced. We propose that the cut-off defining the black hole
entropy should be the dynamically induced length-scale $\bar
\sigma(N)$, \ie, the length scale at which correlations between
configurations on ${\cal J}^+$ and the near horizon region get lost
when $N$ quantum fields contribute to the entropy and therefore to the
near horizon gravitational interactions. {\it We refer the interested reader
to our recent work\cite{Jacobson:2007jx} wherein 
we study the entanglement entropy in the 
presance of dispersion defined in a FF frame such as that of 
eq. (\ref{two}). In that work, it is shown that 
one obtains an entanglement entropy which 
scales as the horizon area per Planck length square
only when the 
$UV$ cutoff (which specifies at which scale measured in the FF frame
the propagation ceases to be the usual Lorentz invariant one) scales as in our 
model, i.e. when the $UV$ cutoff scales as
$1/\bar \sigma \propto M_{Planck}^2/\kappa$.} 

\section*{Acknowledgements} 

I wish to thank D. Arteaga, R.~Balbinot, C. Barrab\`es, R. Brout, L. Ford,
V. Frolov, L. Garay, B.L. Hu, T. Jacobson, S. Liberati, S. Massar,
E. Verdaguer and G.~Volovik for useful discussions.  
The topics here explained have been presented and discussed 
in the following meetings:
the 6-{\it th} and 7-{\it th} Peyresq meetings (France) on {\it
Quantum Spacetime, Brane Cosmology and Stochastic Effective Theories},
the  {\it Artificial Black Holes} meeting held in Rio de Janeiro, 
the 
{\it Black Hole III} conference held in Canada 
and 
the {\it ESF-COSLAB} workshop held in London in July 2001.
I wish to thank the organizers of these meetings.
This work was supported by the NATO Grant CLG.976417.

{\it Finally I am grateful to D. Campo
for a critical reading of the revised manuscript.}

\section{Appendix: The large-$N$ limit}

We briefly mention several interesting features of the large $N$ limit
which illuminate the problems we addressed.

The semi-classical description of quantum processes occurring in a
curved background is based on the following equations
\ba
G_{\mu \nu} &=& 8 \pi G \, \bra{\Psi} T_{\mu \nu} \ket{\Psi} \, ,
\label{GT}\\
\Box_g \hat \phi &=& 0 \, .
\label{Dal}
\ea
In eq. (\ref{Dal}) the field operator propagates in the classical geometry
$g=g_{\mu \nu}^{\Psi}$ which is a solution of eq. (\ref{GT}) driven by the
expectation value $\bra{\Psi} T_{\mu \nu} \ket{\Psi}$ evaluated in the
state $\ket{\Psi}$ using eq. (\ref{Dal}). In this sense, eq. (\ref{GT},
\ref{Dal}) is a self-consistent (Hartree) approximation.

It is quite reasonable that this approximation correctly predicts the
time evolution of {\it certain} quantities in {\it certain}
circumstances, \eg, the {\it rate of mass loss} of a {\it large} black
hole.  However, the criteria which characterize the validity range of
the predictions obtained from eq. (\ref{GT},\ref{Dal}) are not known. An
obstacle in finding these criteria is the identification of the
``small parameter(s)'' which control the deviations between the exact
evolution and 
the semi-classical one 

A rather formal answer to these questions is provided by considering a
large $N$ limit, where $N$ is the number of copies of the $\phi$
field.  The simplest way to understand why the above
equations govern the large $N$ limit is by considering the path
integral approach of matrix elements.  By duplicating $N$ times $\phi$
in eq. (\ref{Z1}) and first integrating over them with a fixed metric $h$,
the $1$-loop effective action for $h$ contains an overall prefactor
$N$ when replacing Newton's constant $G$ by $G/N$. 
Therefore, in a large $N$ limit, the path integral over $h$ can be 
evaluated by a saddle point approximation. 
The stationary phase condition then gives
rise to eq. (\ref{GT}) where the expectation value of $T_{\mu\nu}$
is that of one field. The spread around the saddle point scales like
$N^{-1/2}$.  

In this approach, the validity of the semi-classical 
equations apparantly relies on a statistical argument, as mean quantities 
emerge in the thermodynamic limit. The weakness of this 
argument is
the absence of role played by the hierarchy of the length scales governing the
processes under examination. This is unlike what is found when a 
Born-Oppenheimer treatment is applied to quantum gravity~\cite{Massar:1998cs}.

More interestingly the large $N$ limit makes also predictions {\it
beyond} the semi-classical equations.  For instance, in the limit $N
\to \infty$ with $GN$ fixed, the short distance behaviour of the
graviton propagator is modified, see \cite{Tomb,Verda,HawkS}.  In Minkowski
vacuum, these modifications are of course Lorentz invariant.  However,
in a thermal bath or a curved background, the correction terms will no
longer possess the Lorentz invariant form. Hence they can induce the
effects we are seeking: a dynamically induced scale which breaks the
(local) Lorentz invariance that the un-interacting theory
possessed. Moreover, only low energies (\ie, energies comparable to
the temperature) contribute to this new scale because in the UV
limit all expressions tend to their Minkowski vacuum, and hence
Lorentz-invariant, form.  Hence they shouldn't be any additional
UV divergences
in the expressions giving rise to the new scale, as it is the case for
the corrections to the self-energy of an electron immersed in a
thermal bath.

To further strengthen the relations with what we did in Section 6, it
is instructive to see how the semi-classical treatment emerges from
eq. (\ref{Gmod}) viewed as generating perturbatively the connected Feynman
diagrams when expanding $e^{iS_{int}}$ in powers of $S_{int}$.  
In this description, one finds that the Green function is a
double sum of powers of $N$ and $G$ which possess the following
properties.
\begin{itemize}

\item
The power of $N$ is equal or inferior to that of $G$.

\item
The semi-classical treatment corresponds to the leading series: the
set of graphs weighted by $(GN)^n$. All graphs are
one-particle reducible and are governed by the 
one-point function $\bk{T_{\mu\nu}}$. Upon summing this series, one
identically recovers the Green function evaluated in the `mean'
geometry $g_{\mu \nu}^\Psi$, the solution of eq. (\ref{GT}). 

\item
Having summed up the leading series in $(GN)^n$, our treatment 
corresponds to the {\it next} series: the set of graphs weighted 
by $(G^2 \, N)^n$. All graphs are
two-particle reducible and are governed by the connected
two-point function $\bk{T_{\mu\nu} T_{\alpha \beta}}_C$.  
Upon summing this new series, one obtains the Green function evaluated
in the stochastic ensemble governed by eq. (\ref{mf2}). 
This second series should also be related to the use of the above mentioned 
large-$N$ modified graviton propagator in the place of the unperturbed 
one. 
 
\end{itemize} 
In brief, $N$ is a parameter which organizes the double sum of graphs
into a sum of non-perturbative series. 
When the former series have been summed up, the $m$-th series
contains all powers of $(G^m\,N)$, and is governed by the $m$-th
correlation function of $T_{\mu \nu}$.

The physical question raised by these results is the following: given
the dimensionality of $G=l_{Planck}^2$, can one infer that high orders
in $m$ (the relative power of $G$ with respect to that of $N$) 
become relevant only for high (Planckian) energies ?  We
conjecture that this is the case: the sorting out of graphs in terms
of $m$ is effectively an expansion in the energy of the processes
involved in the matrix element under consideration.  This is what
seems to emerge from our analysis. 
As indicated by eqs. (\ref{18}, \ref{res}), the semiclassical
description of the correlations breaks down when 
$r-2M\simeq \bar \sigma $, \ie,  when the FF 
energy of a mode $\Omega \simeq p_r= \lambda / (r/2M-1) $, see 
eq. (\ref{pr}), reaches the new
scale $1/\bar \sigma$.\footnote{Notice that $\bar \sigma$ 
scales as $N^{-1/2} (GN/M)$.
This writing expresses that in the large $N$ limit at fixed $GN$,
\ie, with a $N$-independent Hawking flux, $\bar \sigma \to 0$ like
$N^{-1/2}$, thereby verifying that in this limit the scale which
signals the breakdown of the semi-classical description indeed
vanishes.}  We thus find, as in \cite{Massar:1998cs},
that the validity of the semi-classical equations relies on a {\it
hierarchy} of energy scales. Indeed, for a large black hole, 
$\bar \sigma \ll M$ even when $N=1$. Thus $N$ is 
{not necessary} to justify the validity of the semi-classical 
description. It is rather a useful parameter which helps sorting out 
the different contributions in radiative corrections.
In particular, it allows to discard the self-interacting (non-planar)
graphs which are more difficult to evaluate (since they 
cannot be expressed in terms of metric fluctuations in the present case)
because their power of $N$ is smaller those weighted by $(G^m\,\! N)^n$.


\end{document}